\documentclass[pre,aps]{revtex4}
\usepackage{graphicx}
\usepackage[T2A]{fontenc}\selectfont
\usepackage{amsopn}
\usepackage{amsfonts,bm}
\usepackage{amsmath}
\usepackage{latexsym}

\begin{document}
\large

\title{Kinematic dynamo in two-dimensional chaotic flow: the initial and final stages.}

\author{I.V.Kolokolov}

\affiliation{Landau Institute for Theoretical Physics, RAS, \\
 119334, Kosygina 2, Moscow, Russia}

\date{\today}

\begin{abstract}
The small-scale kinematic dynamo in a two-dimensional chaotic flow is studied. 
The analytic approach is developed in framework of the Kraichnan-Kazantsev model. 
It is shown that the growth of  magnetic field $\bm{B}$ fluctuations stops at large times in accordance with 
so-called anti-dynamo theorems. The value of  $\bm{B}^2$ increased therewith in square of the magnetic Prandtl 
number times. The spatial structure of the correlation tensor of the magnetic field is found.
\end{abstract}

\maketitle

\section{Introduction}

The dynamo phenomenon consists in amplification of magnetic field in a conducting fluid if it was present there in an initial time moment \cite{M78,P79,ZRS,CG95,LL}. 
If the fluid has infinite conductivity then this amplification is a direct consequence of the magnetic flux conservation for any liquid contour. 
Indeed, for a generic fluid flow one can find in vicinity of a given point some liquid contour with diminishing square. Conservation of the flux of the magnetic field 
  $\bm{B}$  through  this contour is possible only if  $|\bm{B}|$ increases. Thus, in the limit of zero resistivity the effect is local and weakly depending on 
large-scale properties of the flow. If the resistance of the fluid  is small but finite the dissipation will govern large time behavior of the megnetic field distribution.
It does not mean that the values of  $\bm{B}^2, \bm{B}^4, \dots$  averaged over space will cease growing. Rather, global characteristics of the flow becomes important.

The evolution equation of the field  $\bm{B}({\bm r}, t)$ in the incompressible flow ${\bm v}({\bm r}, t)$ has the form \cite{LL}:
 \begin{equation}
 \partial_t{\bm B}=
 ({\bm B}\cdot{\bm\nabla}){\bm v}
 -({\bm v}\cdot{\bm\nabla}){\bm B}
 + \kappa\nabla^2{\bm B}.
 \label{e11}
 \end{equation}
where $\kappa$ is the magnetic diffusion  coefficient inversely proportional to the fluid conductivity. 
We study here the kinematic regime when the back reaction of the magnetic field on the flow can be neglected. The local properties of the velocity field are determined 
by the viscous scale $R$. In the vicinity $|{\bm r}-{\bm r}(t)|\ll R$ of a given Lagrangian trajectory ${\bm r}(t)$ the velocity can be approximated by a linear profile:
\begin{equation}
 v_\mu(\bm{r},t)\approx V_\mu^{(0)}(t)+ \sigma_{\mu\nu}(t)r_\nu, \quad V_\mu^{(0)}(t)=v_\mu(\bm{r}(t),t),\quad \sigma_{\mu\mu}=0.
 \label{linprof}
 \end{equation}

The magnetic diffusion is significant on the scales less then $r_d\sim\sqrt{\kappa/\lambda}$ where $\lambda\sim |\hat{\sigma}|$ is the characteristic Lyapunov exponent 
of divergence of close Lagrangian trajectories in the flow. We consider here the case when the ratio $R/r_d$ called the magnetic Prandtl number $Pr_m$ is large:
$ Pr_m=R/r_d\gg 1$  \cite{Sch,00GS,BSS}. If the linear approximation (\ref{linprof}) is valid the advection  $\bm{V}^{(0)}$ can be excluded be the Galilean transformation and the 
evolution equation for the spatial Fourier components of the magnetic field 
\begin{equation}
 B_\alpha(\bm{r},t)=\int\frac{d^2{\bf k}}{(2\pi)^2}e^{i{\bf k}{\bf r}}B^\alpha_{\bm k}(t)
 \label{fur}
 \end{equation}
becomes the first-order partial differential equation:
\begin{equation}
\partial_t B^\alpha_{\bm k} -
\sigma_{\mu\nu}k_\mu\frac{\partial}{\partial k_\nu} B^\alpha_{\bm k}-B^\nu_{\bm k}\sigma_{\alpha\nu}+\kappa {\bm k}^2 B^\alpha_{\bm k} =0. 
 \label{lpfur}
 \end{equation} 
It can be solved both for random (see \cite{CFKV,01FGV}) and regular matrix $\hat{\sigma}(t)$. Let us consider the simplest version of this local evolution when $\hat{\sigma} $ 
is the diagonal matrix constant in time:
\begin{equation}
\sigma_{\mu\nu}(t)=\rm{diag}(\lambda, -\lambda, 0).
 \label{linprof1}
 \end{equation}
As an illustrative example let us take the Gaussian profile for the initial field distribution:
 \begin{equation}
 {\bm B}( {\bm r},t=0)=\text{curl} {\bm A},\quad
 {\bm A}={\bm a}\exp\left(-\frac{{\bm r}^2}{4L^2}\right),{\bm a}=(0,a,0).
 \label{sig}
 \end{equation}
The solution of the initial data problem in the dissipative regime  $r_d e^{\lambda t}\gg L$ has the form:
 \begin{eqnarray}
 && B_1\approx \frac{a r_3}{2Lr_d\left(1+\kappa t\right/L^2)^{3/2}}
 \exp\left\{-\frac{r_3^2}{4(L^2+\kappa t)}-\frac{r_2^2}{4r_d^2}-\frac{r_1^2}{4L^2}e^{-2\lambda t}\right\}, 
 \label{strs}\\
&& B_3 \approx -r_1 \frac{a e^{-2\lambda t}}{2L r_d \left(1+\kappa t\right/L^2)^{1/2}}
 \exp\left\{-\frac{r_3^2}{4(L^2+\kappa t)} -\frac{r_2^2}{4r_d^2} -\frac{r_1^2}{4L^2}e^{-2\lambda t} \right\}.
\label{thco}
 \end{eqnarray}
For the localized blob (\ref{sig}) the exponential growth of the field amplitude stops at dissipative times. However, the exponential decay of the component $B_3$ does not 
prevent $B_1$ from being (quasi) frozen despite the solenoidality:  $\partial_1 B_1+\partial_3 B_3=0$. That is, the field component along the stretching  direction is flattened out by the flow 
slowly changing its amplitude. Let us take now the field blobs randomly distributed over space to be the initial data for the Eq.  (\ref{lpfur}):
\begin{equation}
 {\bm B}({\bm r},0)=\text{curl} \left[\sum_{\bm n}{\bm a}_{\bm n}\exp\left(-\frac{({\bm r}-{\bm R}_{\bf n})^2}{4L}\right)\right],\quad
{\bm a}_{\bm n}=(0,a_{\bm n},0), {\bm n}=(n_1,n_2,n_3).
 \label{ig}
 \end{equation}
The  amplitude $B_1({\bm r},t)$  at a given time moment  $t\gg\lambda^{-1}\ln(L/r_d)$   can be represented  as a sum 
of exponentially large number of addends:
\begin{equation}
B_1\propto \sum\limits_{-M}^{M}a_{(0,m,0)},\qquad M\sim r_de^{\lambda t}/L.
\label{sum}
 \end{equation}
If $a_{(0,m,0)}$ are uncorrelated random numbers then we arrive to the estimation  $B_1\propto e^{\lambda t/2}$.
The exponential growth of the magnetic field fluctuations continues due to aggregation of the initial blobs in the contracting direction of the flow. 
The magnetic diffusion homogenizes the field on the scales $\sim r_d$. For random matrix $\hat{\sigma}(t)$  the Lyapunov exponent $\lambda$ is a fluctuating quantity. The moments 
of the field   ${\bm B}$ can be evaluated with the same estimation $|{\bm B}|\propto e^{\lambda t/2}$ averaged 
over the statistics of $\lambda$:  $\langle {\bm B}^{2n}(t) \rangle \propto \langle \exp(n\lambda t) \rangle $. The resulting statistics of the magnetic field is intermittent 
because of the inequality  $\langle \exp(n\lambda t) \rangle \gg  \exp(n\langle\lambda \rangle t)$ for $\lambda t \gg 1$  \cite{CFKV, 01FGV,kogan}. 

The key feature of the example providing the field enhancement is the absence of (anti) correlations between field amplitudes in different blobs delivered by the  flow 
along the contracting direction 
to the observation  point. But for a finite correlation length $R$ of the velocity field this property may cease to hold for large enough $\lambda t$. Indeed, in the course of evolution 
the magnetic lines of force form narrow strip-like clusters with the widths  $\sim r_d$  and exponentially growing lengths. The field ${\bm B}$ is strongly correlated along such strips 
due to the flattening out effect mentioned above. The correlation length $R$ of the velocity field turns out to be the characteristic curvature radius of these strips. As a result the magnetic
field  becomes correlated  along random curves in the plane or in the space. Hence, several parts of the same strip fall within the contracting domain of the flow. The 
anticorrelation arising in this way with certain probability can either modify the growth of the field moments or stop it at all. We show here that the latter possibility 
realises in two-dimensional random flow. One can note some similarity of the phenomenon with weak localization \cite{LarKhm}. 

In the present paper we study the dynamo in two-dimensional chaotic flow. Such liquid motion can be realized directly \cite{Shats} and  this description  can be also be 
applied strongly rotating hydrodynamical systems  \cite{eink, gui}.
The velocity field is supposed to have the finite correlation length $R$. This formulation of the problem 
is the subject of so-called '' antidynamo theorems '' going back to Zeldovich's  works \cite{Z56, ZR}, see also the recent review \cite{so14}. 
In the papers it is noted correctly that the transverse component $B_3$ of the magnetic field behaves like a decaying passive scalar field. The remaining  two components should decay in the 
same manner according to  \cite{Z56, ZR}. The simple example  (\ref{linprof1})-(\ref{sum}) demonstrates the fallacy of this argumentation. On the other hand, there are rigorous 
assertions \cite{Ose,Arno} claiming the absence of unlimited growth of the magnetic field in this geometry. These theorems relay to flows on compact manifolds what 
seems to be tantamount to the finite correlation length of the randon velocity field. The possibility of an initial growth of the field is pointed out in mentioned papers, 
although the appropriate estimations of times and amplitudes are absent. 

It is shown in the present paper that for  $\lambda t\gg \ln R/r_d$ the field moments stop to grow in accordance  with  anti-dynamo statements. However, the value of 
$\langle {\bm B}^2\rangle$ increases in the course of evolution in  $\sim (R/r_d)^2$  times. This factor can reach the order of magnitude of $\sim 10^{28}$ \cite{Sch} and 
it can switch on non-linear effects \cite{Sche2}.
 
\section{The Kraichnan-Kazantsev model in two dimensions.}

The analytic description of the phenomenon is developed here for the Kraichnan-Kazantsev model 
\cite{Kra,Kaz} where the velocity ${\bm v}({\bm r},t)$ statistics is supposed to be Gaussian with zero mean value and 
the pair correlator of the form: 
\begin{equation}
\langle v_{\mu}(\bm{r},t)v_{\nu}(\bm{r'},t')\rangle=
 \delta(t-t'){\cal C}_{\mu\nu}(\bm{r}-\bm{r'}).
 \label{velt1}
 \end{equation}
We consider the initial magnetic fiel to be randomly distributed over the space. The evolution equation for the correlation tensor 
$F_{\alpha\beta}({\bm r},t)=\langle B_\alpha ({\bm r}',t) B_\beta ({\bm r}'+{\bm r}, t)\rangle$ follows directly from the equation of motion (\ref{e11}):
 \begin{equation}
 \partial_t F_{\alpha\beta}=\left[{\cal C}_{\mu\nu}(0)-{\cal C}_{\mu\nu}(\bm{r})\right]\partial_\mu\partial_\nu F_{\alpha\beta}+
\partial_\mu {\cal C}_{\nu\beta}(\bm{r})\partial_\nu  F_{\alpha\mu}+\partial_\mu {\cal C}_{\nu\alpha}(\bm{r})\partial_\nu  F_{\mu\beta}-
 F_{\mu\nu}\partial_\mu\partial_\nu {\cal C}_{\alpha\beta}(\bm{r})+2\kappa\triangle  F_{\alpha\beta}.
 \label{ceq1}
 \end{equation}
The magnetic field has all three components depending on the third coordinate $r_3$ as well. On the other hand, the tensor ${\cal C}_{\mu\nu}$ does not depend on  $r_3$ 
and the corresponding components of ${\cal C}_{\mu\nu}$ are equal to zero:  ${\cal C}_{\mu 3}= {\cal C}_{3\nu}=0.$   In this case there is a closed evolution equation for in-plane components 
$ F_{\alpha\beta},\alpha, \beta=1,2$ which has the form  (\ref{ceq1}) where the indexes take the values 1 and 2. We deal below with this components only. 
The in-plane reduced correlation tensor has non-zero divergency:   $ \partial_\alpha F_{\alpha\beta} \neq 0$. The coordinate   $r_3$ is a parameter in the two-dimensional problem and 
it is not pointed out in the sequel explicitly. 

We consider the statistics of the velocity field to be isotropic:
\begin{equation}
{\cal C}_{\mu\nu}(\bm{r})=\delta_{\mu\nu}{\cal C}_1(r)+\frac{r_\mu r_\nu}{r^2}{\cal C}_2(r).
 \label{velt2}
 \end{equation}
The incompressibility of the flow leads to the following relation between ${\cal C}_1(r)$ and ${\cal C}_2(r)$:
 \begin{equation}
 {\cal C}_1^\prime(r)=- {\cal C}_2^\prime(r)-\frac{1}{r}{\cal C}_2(r),\quad
 {\cal C}_1(r)=V_0 R- {\cal C}_2(r)-\int\limits_0^r\,\frac{dr'}{r'} {\cal C}_2(r'),\quad {\cal C}_2(0)=0.
 \label{vee1}
 \end{equation}
The value $V_0$ has the sense of the typical advection velocity and is defined by the condition 
${\cal C}_1(r\to\infty)={\cal C}_2(r\to\infty)=0$. The statistical isotropy of the magnetic field leads to the decomposition of the tensor 
$F_{\alpha\beta}({\bm r},t)$ similar to (\ref{velt2}):
\begin{equation}
F_{\alpha\beta}({\bm r},t)=\delta_{\alpha\beta}\mathcal S(r,t)+\frac{r_\alpha r_\beta}{r^2}\mathcal Y(r,t).
 \label{mfi1}
 \end{equation}
One can check that the evolution of the function 
\begin{equation}
 \Phi(r,t)=(1+r\partial_r)\mathcal Y(r,t)+r\partial_r \mathcal S(r,t),\quad 
\partial_\alpha F_{\alpha\beta}=\frac{r_\alpha}{r^2}\Phi(r,t),
 \label{divo}
 \end{equation} 
decouples:   
\begin{eqnarray}
\label{uedfi1}
\partial_t \Phi=
\left[{\cal C}_1(0)-{\cal C}_1(r)-{\cal C}_2(r)\right]\left(\partial_r^2 \Phi- \frac{1}{r}\partial_r \Phi\right)
+r_d^2\left(\partial_r^2 \Phi-r^{-1}\partial_r \Phi\right)  .
\end{eqnarray}
When such is the case the growth of the magnetic field is determined by the dynamics at the scales  $r\ll R$ where one can use the expansion:
\begin{equation}
\langle v_{\mu}(\bm{r},t)v_{\nu}(\bm{0},t')\rangle \approx \left[V_0 R\delta_{\mu\nu}-
\lambda\left(3r^2\delta_{\mu\nu}-2r_\mu r_\nu\right)\right]
\delta(t-t').
 \label{velt}
 \end{equation}
The Lyapunov exponent  $\lambda$ can be expressed as $\lambda= V_0/R$. If (\ref{velt}) works the equations for the functions $\Phi(r,t)$ and $Y(r,t)$ have the form:
\begin{eqnarray}
\label{uedfi}
\lambda^{-1}\partial_t \Phi= \hat{\mathcal L}_\Phi \Phi, \quad 
\hat{\mathcal L}_\Phi  =
r^2\partial_r^2 - r\partial_r +
r_d^2\left(\partial_r^2 -r^{-1}\partial_r \right),
\end{eqnarray} 
\begin{eqnarray}
\label{uedd}
\lambda^{-1}\partial_t \mathcal Y= \hat{\mathcal L}_{\mathcal Y}\mathcal Y -8\Phi,\quad 
\hat{\mathcal L}_{\mathcal Y}=r^2\partial_r^2 +7r\partial_r +8+
r_d^2\left(\partial_r^2 +r^{-1}\partial_r -4r^{-2}\right).
\end{eqnarray}
The dissipative scale $r_d=\sqrt{2\kappa/\lambda}$ is considered to be smallest in the problem. At largest distances  $r\gg R$ the correlation tensor obey
the diffusion equation:
\begin{equation}
\partial_t F_{\alpha\beta}=\eta\triangle  F_{\alpha\beta}, \quad \eta=\lambda R^2.
 \label{fid}
 \end{equation}   

We intend to find leading asymptotics of the functions $\Phi(r,t)$, $\mathcal  S(r,t)$ and $\mathcal Y(r,t)$ at $\lambda t\gg 1$ and $r\gg r_d$ up to numerical factors.
Notice that there is some variety in asymptotical regimes because of the existence of the large ratio $R/r$ if $r\ll R$. We present derivation of $\Phi(r,t)$ and $\mathcal Y(r,t)$; 
the function $\mathcal S(r,t)$ can be restored easily. 

To solve the evolution equations we use the Laplace transform:
\begin{equation}
\mathcal Y_p(r)=\lambda\int\limits_0^\infty dt\, e^{-p\lambda t}\mathcal Y(r,t),
 \label{lt}
 \end{equation}   
The function $\Phi_p(r)$ defined analogously is equal to the convolution of the initial data $\Phi^{(0)}(r)=\Phi(r,t=0)$ with the Green function:
\begin{equation}
\Phi_p(r)=\int\limits_0^\infty dr^\prime\, G_\Phi(p|r,r^\prime)\Phi^{(0)}(r^\prime),
 \label{fg1}
 \end{equation}   
The dissipative part in (\ref{ceq1}), (\ref{uedfi}) is inessential at  $r\gg r_d$ and the length  $r_d$  is used below as a lower cut-off in  
$dr^\prime$-integration if it is required. For  $r\geq r^\prime, r^\prime \ll R$ the Green function $G_\Phi(p|r,r^\prime)$  has the form:
\begin{equation}
G_\Phi(p|r,r^\prime)= \varphi_p(r)\frac{\left(r^\prime\right)^{-2+\sqrt{p+1}}}{2\sqrt{p+1}},
 \label{fg3}
 \end{equation} 
where $\varphi_p(r)$ at $r\ll R$ is a linear superposition of power functions:
\begin{equation}
\varphi_p(r)=r^{1-\sqrt{p+1}}+b_p R^{-2\sqrt{p+1}} r^{1+\sqrt{p+1}}, \quad r\ll R,
 \label{fifu}
 \end{equation} 
with the coefficient $b_p$ to be determined by matching with the domain $r\geq R$. 
For $r\gg R$ and $Re\, p>0$ the function  $\varphi_p(r)$  is a decaying solution of the equation:
\begin{eqnarray}
\label{phid}
\left(p/R^2-\partial_r^2 +r^{-1}\partial_r \right)\varphi_p(r)=0,
\end{eqnarray}
and has the form
\begin{equation}
\varphi_p(r)\approx \mathcal{B}_p R^{-\sqrt{p+1}}r K_1\left(\frac{r}{R}\sqrt{p}\right), \quad r\gg R.
 \label{fid1}
 \end{equation}  
The solution in the domain $r\sim R$ defines $2\times 2$-matrix    $\hat{g}$  matching asymptotics (\ref{fifu}) and  (\ref{fid1}) and fixing the coefficients $\mathcal{B}_p$ and $b_p$:
\begin{equation}
\left(b_p+1, b_p+1+\sqrt{p+1}(b_p-1)\right)\hat{g}=\mathcal{B}_p\left(K_1(\sqrt{p}), -\sqrt{p}K_0(\sqrt{p})\right).
 \label{gma}
 \end{equation}
For regular functions    ${\cal C}_{1,2}(r)$ the elements of $\hat{g}$ have no singular points in the $p$-plane. Thence the singularities of  $b_p$ and $\mathcal{B}_p$ determining 
large time behavior emerge from the right hand side of  (\ref{gma}) and are placed at  $p=0$. The leading terms in expansion of $b_p$ near $p=0$ are 
\begin{equation}
p\to 0, \quad b_p\to b_0+b_1 p\ln p, \quad b_{0,1}\sim 1.
 \label{bpz}
 \end{equation} 
If  $r\gg R$  and $p\to 0$ the Laplace transform $\varphi_p(r)$  is proportional to the modified Bessel function:
\begin{equation}
p\to 0, \quad \varphi_p \propto f_0\frac{r}{R} p K_1\left(\frac{r}{R}\sqrt{p}\right).
 \label{lds}
 \end{equation}   
The Eq.(\ref{uedd}) for the function  ${ \mathcal Y}(r,t)$ at $r\ll R$ has the source $-8\Phi(r,t)$. Thus the Laplace transform 
$\mathcal Y_p(r)$ is expressed both in terms of the initial condition 
${ \mathcal Y}^{(0)}(r)={ \mathcal Y}(r,t=0)$ and  $\Phi_p(r)$:
\begin{equation}
\mathcal Y_p(r)\approx\int\limits_0^R dr^\prime\, G_{ \mathcal Y}(p|r,r^\prime)\left({ \mathcal Y}^{(0)}(r^\prime)-\Phi_p(r^\prime)\right),
 \label{fgy1}
 \end{equation}   
where $ G_{ \mathcal Y}(p|r,r^\prime)$ for $r^\prime >r$ and  $r,r^\prime \ll R$ has the form:
\begin{equation}
G_{ \mathcal Y}(p|r,r^\prime)= 
r^{-3+\sqrt{p+1}}\,
\frac{\left(r^\prime\right)^{2-\sqrt{p+1}}}{2\sqrt{p+1}}\left[1+a_p\left(\frac{r^\prime}{R}\right)^{2\sqrt{p+1}}\right].
 \label{fgy2}
 \end{equation} 
The coefficient $a_p$ is defined like $b_p$ by the matching with the domain $r\geq R$ and has the logarithmic branching point at $p=0$:
\begin{equation}
p\to 0, \quad a_p\to a_0+a_1 p^2\ln p, \quad a_{0,1}\sim 1.
 \label{apz}
 \end{equation} 

For $p\to 0$ and  $r\gg R$ the expression for ${ \mathcal Y}_p(r)$ follows from (\ref{fid}):
\begin{equation}
p\to 0, \quad  { \mathcal Y}_p(r)\propto f_0 pK_2\left(\frac{r}{R}\sqrt{p}\right).
 \label{lds1}
 \end{equation}   
Performing  the inverse Laplace transform to restore the functions  $\Phi(r,t)$ and ${ \mathcal Y}(r,t)$ the inequality  $\lambda t \gg \ln L/r_d$ is assumed. 
It corresponds to the dissipative stage of the dynamics. We are interesting primarily in magnetic field fluctiations on the scales $r\ll R$. Although the local 
expansion (\ref{velt}) the field dynamics is strongly affected by the finiteness  of R with time as we see below.

There are two asymptotical regimes in the evolution of the function  $\Phi(r,t)$. During the first one going on at $2\lambda t<\ln R/r_d$ the shape of $\Phi(r,t)$ 
may be described as an exponentially blowing hull:
\begin{equation}
\Phi(r,t)\propto \frac{f_0}{\sqrt{\lambda t}} \frac{r}{r_d}\exp\left(-\lambda t - \frac{1}{4\lambda t}\ln^2  \frac{r}{r_d}\right)
 \label{infla1}
 \end{equation} 
One can see that   $\Phi(r,t)$  is concentrated in a  narrow neighborhood of the running point   $r_m(t)=r_d\exp(2\lambda t)$; the value  $\Phi_m=\Phi(r_m(t),t)$ decreases 
slowly: $\Phi_m \sim f_0(\lambda t)^{-1/2}$. When $r_m(t)$ reaches the velocity scale $R$ the behavior of  $\Phi(r,t)$  changes:
\begin{equation}
\Phi(r,t)\propto f_0 \frac{r^2}{R^2}\frac{1}{(\lambda t)^2}, \quad
\lambda t\gg \ln R/r\gg 1. 
 \label{infla2}
 \end{equation}  
Restoring the function ${ \mathcal Y}(r,t)$ note that the Green function (\ref{fgy2}) has the branching point at $p=-1$ together with the singularity at  $p=0$.
This leads to the monotonic decay ($\sim \exp(-\lambda t)$ at the first stage of the evolution) of the initial data contribution which  is ignored therefore below.
On the other hand the kernel  (\ref{fgy2}) grows with  $r^\prime$ at  $p<3$, $\operatorname{Im} \, p=0$. In its turn the source  $\Phi_p(r)$ grows with  $r$ and as a result 
 ${ \mathcal Y}(r,t)$ increases exponentially at intermediate times:
\begin{equation}
{ \mathcal Y}(r,t)\propto f_0 \frac{L}{r} \exp(3\lambda t), \quad \ln \frac{L}{r_d}\ll \lambda t \leq \frac{1}{4}\ln \frac{R^2}{L r}.
 \label{ydy1}
 \end{equation}  
In the next time inerval the grows of ${ \mathcal Y}(r,t)$ continues but it becomes $R$ - dependent:
\begin{equation}
{ \mathcal Y}(r,t)\propto \frac{f_0}{\sqrt{\lambda t}} \frac{R^4}{r^3 L}\exp\left(-\lambda t - \frac{1}{4\lambda t}\ln^2  \frac{R^2}{r L}\right),
\quad \frac{1}{4}\ln \frac{R^2}{L r}\leq \lambda t \leq \frac{1}{2}\ln \frac{R^2}{L r},
 \label{ydy2}
 \end{equation} 
The maximal value reached by ${ \mathcal Y}(r,t)$ is parametrically large:  ${ \mathcal Y}_{max}(r)\sim R^2/r^2$. The behavior of  ${ \mathcal Y}(r,t)$ at $\lambda t\gg \ln(R/r)$ 
is governed by the singularity of ${ \mathcal Y}_p(r)$ at $p=0$:
\begin{equation}
{ \mathcal Y}(r,t)\propto f_0 \frac{R^2}{r^2}\frac{1}{(\lambda t)^2}, \quad \lambda t\gg \ln \frac{R^2}{L r}.
 \label{yd3}
 \end{equation}  
The complete
correlation tensor at large times and  $r\ll R$ 
is restored noting that $\Phi$ can be neglected it the relation (\ref{divo}):
\begin{equation}
F_{\alpha\beta}(r,t)\propto \frac{f_0}{(\lambda t)^2}\left(-\delta_{\alpha\beta} + 2\frac{r_\alpha r_\beta}{r^2}\right) \frac{R^2}{r^2},\quad 
r_d\ll r\ll R,\quad F_{\alpha\beta}(r,t)\propto \frac{f_0}{(\lambda t)^2} \frac{R^2}{r_d^2}\delta_{\alpha\beta}, \quad r\alt r_d. 
 \label{fcomp}
 \end{equation}  

\section{Conclusion.}
We see that in two-dimensional chaotic flow the fluctuations of the magnetic field increase their amplitude in $R/r_d$ times. The statistics of the field in the course 
of the exponential growth is highly intermittent similar to the three-dimensional case \cite{CFKV}: the field energy is concentrated in strip-like domains with widths of the 
order of $r_d$. After $t\sim \lambda^{-1}\ln(R/r_d)$ the growth gives way to the slow decrease. The formal cause for a stop of the dynamo is the fictitious character of the 
singularity of  ${ \mathcal Y}_p(r)$ at $p=3$:
\[
{ \mathcal Y}_p(r)\propto \frac{1}{4-2\sqrt{p+1}}\left[1-\left(\frac{r}{R}\right)^{4-2 \sqrt{p+1}}\right],
\]
This expression has the pole at  $p=3$ in the limit $R\to\infty$ only. For any finite  $R$ there is no true singularity at $p=3$ and the exponential growth is an intermediate 
asymptotics. It is instructive to compare the situation with three-dimensional model  \cite{Kaz}. In this case the closed equation for the trace ${ \mathcal F}=F_{\alpha\alpha}$ 
can be derived:
\begin{eqnarray}
\label{3d1}
\lambda^{-1}\partial_t \mathcal F= \hat{\mathcal L}_{\mathcal F}\mathcal F, \quad 
\hat{\mathcal L}_{\mathcal F}=r^2\partial_r^2 +6r\partial_r +10, \quad r_d\ll r\ll R.
\end{eqnarray} 
It can be checked easily that the Laplace transform  ${\mathcal F}_p(r)$  has the true singularity at  $p=15/4$:
\[
{\mathcal F}_p(r)\propto f_0 \left(\frac{L}{r}\right)^{5/2}\frac{1}{\sqrt{p-15/4}}
\left[1- \left(\frac{r}{R}\right)^{\sqrt{p-15/4}}\right].
\]
The cancellation of the divergence does not lead to elimination of the branching point at  $p=15/4$. Thus the global structure of the flow affects the dynamo weakly; 
this is in agreement with the results of   \cite{Vince} and justified the local approach used in \cite{CFKV,01FGV} to find the multipoint correlation functions of the magnetic field.
In two-dimensional case it can be done in the initial stage   $\lambda t \ll \ln (R/r_d)$ only. 

If the volume  $\upsilon_0$ occupied by the initial field fluctuations is finite the exponential growth of their amplitude is an intermediate asymptotics for 
$t\ll \lambda^{-1}\ln (\upsilon_0/r_d^3)$ even in three-dimensional case.

I am grateful to V.V.Lebedev for numerous discussions and stimulating questions. I wish to thank E.A.Kuznetsov, A.I.Milstein and G.E.Falkovich for helpful notices and benevolent criticism. 
The work is supported by RScF grant 14-22-00259.

\end{document}